# Anomalous Topological Bloch Oscillations under Non-Abelian Gauge Fields

Chunyan Li,[1,*] Ce Shang,[2,†] and Boris A. Malomed[3]

[1]*School of Physics, Xidian University, Xi'an 710071, China*

[2]*Chinese Academy of Sciences, Aerospace Information Research Institute, 100094, Beijing, China*

[3]*Instituto de Alta Investigacio´n, Universidad de Tarapac´a, Casilla 7D, Arica, Chile*

Topological Bloch oscillations are a hallmark of quantum transport phenomenon in which wavepackets undergo oscillatory motion driven by the interplay between an external force and topological edge states and serve as a powerful dynamical probe for the geometric properties of topological bands. Spin-orbit coupling (SOC) has also emerged as a crucial ingredient for manipulating quantum states in materials, with the corresponding gauge fields arising from the Rashba and Dresselhaus interactions. In this work, we investigate the propagation of spinor wavepackets in a honeycomb Zeeman lattice governed by the Gross-Pitaevskii equation. By tuning the relative strengths of Rashba and Dresselhaus SOC, we engineer a non-Abelian gauge field that drives anomalous topological Bloch oscillations (ATBOs). Unlike conventional topological Bloch oscillation (TBOs), these ATBOs exhibit asymmetric motion, including a freezing effect in one half of the oscillation cycle, which can be tuned by the SOC parameters and external forces. Our findings establish SOC-based non-Abelian gauge fields as a powerful mechanism controlling topological quantum dynamics, with implications for spintronic devices and quantum data processing.

## Introduction

Bloch oscillations (BO) represent a fascinating quantum phenomenon in condensed matter, with electrons in periodic potentials performing oscillatory motion, rather than linear acceleration, under the action of a gradient force. First predicted by Bloch and Zener in the early days of quantum mechanics [1, 2], these oscillations arise from the periodicity of crystal lattices and the corresponding band structures, which demonstrate the interplay between topological phases of matter and non-equilibrium quantum dynamics. For decades, Bloch oscillations remained a theoretical curiosity due to the challenges posed by scattering processes in natural crystals, until their experimental observation in semiconductor superlattices [3–7], and later in ultracold atomic gases [8–11] and photonic structures [12–26]. Under crossed fields, BOs in inversion-broken monolayer phosphorene show confined–deconfined phase transitions [27], while in Rashba spin-orbit-coupled (SOC) systems, it can be modified by hexagonal warping [28]. Additionally, the research extended to Fibonacci anyons correlates BOs dynamics with non-Abelian gauge field physics [29].

Parallel to these developments, the advent of topological materials has revolutionized the understanding of quantum phases of matter [30–32]. The discovery of topological insulators [33, 34], topological semimetals [35, 36], and higher-order topological insulators [37] has revealed a rich landscape of quantum states characterized by nontrivial bulk invariants and topologically protected boundary states. In these settings, the Berry curvature, i.e., a gauge field in the momentum space, plays a fundamental role in determining electron dynamics, giving rise to anomalous velocity terms that significantly modify the semiclassical transport [38]. Meanwhile, various forms of SOC [39, 40] has emerged as a crucial ingredient for manipulating quantum states in materials [41–44]. In particular, the combination of Rashba and Dresselhaus SOC terms, arising from the structural and bulk inversion asymmetry, respectively, induces effective momentum-dependent magnetic fields that strongly affect electron spin dynamics [45–53]. When these SOC terms coexist with comparable strengths, they can give rise to non-Abelian gauge potentials, in which the components of the effective magnetic field do not commute, leading to rich spin textures and topological phenomena [54].

Moreover, the non-commutative gauge field unlocks transport regimes beyond standard topological insulators, including spacetime-modulated gauge field-driven parametric resonances [55] and non-Abelian band singularities transcending conventional Chern or $Z_2$ invariants [56]. In non-Hermitian systems, it enhances the robustness and self-healing of topological modes [57], underscoring broad applicability in open quantum platforms. In ultracold atomic gases, a synthetic non-Abelian topological interface can be realized by tuning the spatial offset of SOC fields, which provides an experimentally feasible approach to explore non-Abelian topological properties [58]. Very recently, this concept has been extended to photonic synthetic dimensions, where synthetic non-Abelian electric fields coupled with SOC offer a powerful platform for engineering complex gauge structures [59].

The interplay between gradient forces and nontrivial topology gives rise to topological Bloch oscillations (TBOs) [19, 24, 42, 60, 61], which exhibit dynamics fundamentally distinct from their nontopological counterparts. Unlike conventional BOs, where wavepackets are confined to a single band and undergo oscillatory motion, TBOs involve interband mixing, facilitated by edge states. The wavepacket can merge into a chiral edge mode, pass through the topological bandgap, and reappear at the opposite boundary of the topological setup. This mechanism results in large-scale real-space displacements of boundary states, allowing TBOs to serve as dynamical probes for topological invariants and the bulk-boundary correspondence [19, 24, 42, 62]. Subsequent studies extended this concept to higher-order topological insulators, where Bloch oscillations emerge due to the interplay between the non-Abelian Berry curvature of degenerate band structures and quantized Wilson loops [60]. In three-dimensional settings, engineered non-Abelian couplings have been shown to induce higher-order topological states accompanied by directional Bloch oscillations [63]. Furthermore, Bloch oscillations have been explored in diverse topological platforms, including Weyl semimetal slabs where surface Fermi arcs and bulk states exhibit distinct oscillatory behaviors [64], in HgTe quantum wells showcase BOs tied to their topological phase diagram [65].

Aside from these advances, the interplay between non-Abelian SOC and topological Bloch oscillations needs to be further explored. Understanding how these non-commuting fields steer wavepackets could provide the physical foundation for advanced spin-logic and valleytronic devices. In this work, we aim to bridge this gap by analyzing topological BO in systems with an engineered Rashba–Dresselhaus SOC. We demonstrate that the SOC's non-Abelian nature steers anomalous topological Bloch oscillations (ATBOs), providing precise control over the asymmetric oscillatory dynamics, including a tunable-"freezing" effect during half of the cycle, which is a crucial departure from the familiar symmetric TBOs. Our results suggest a new setup for controlling quantum dynamics in topological systems, with feasible applications to spin-based electronics.

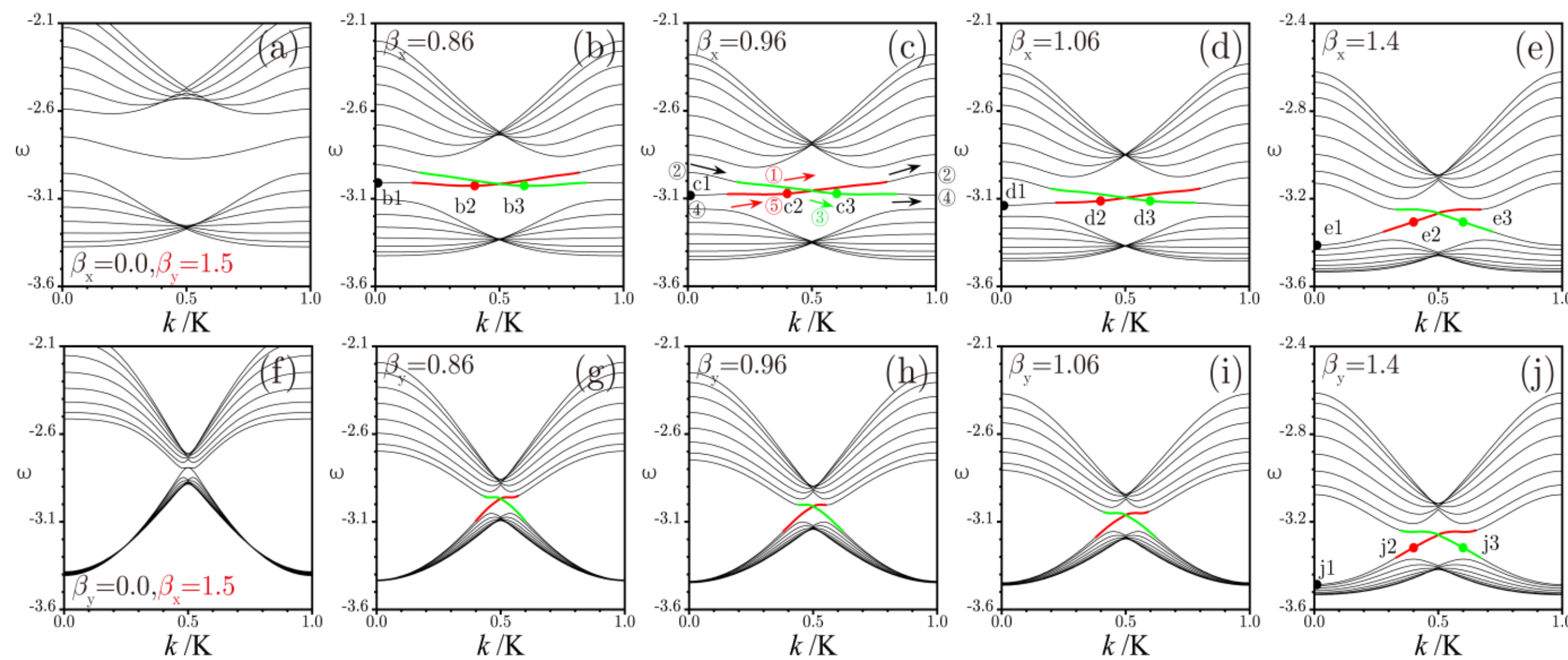

**Fig. 1** Energy-momentum diagrams for SOC with unequal Rashba and Dresselhaus ingredients, *viz*, $\beta_x$= 0.0, 0.86, 0.96, 1.06, 1.4 at $\beta_y$= 1.5 (a-e) and $\beta_y$= 0.0, 0.86, 0.96, 1.06, 1.4 at $\beta_x$= 1.5 (f-j). The colored curves correspond to edge states, while the modes denoted by solid circles are presented in **Fig. 2**. Colored arrows alongside with circled number guide the move path across the momentum space at five specific time $t = 0, T/4, T/2, 3T/4, T$

## The model: spinor Gross-Pitaevskii equations

The action of SOC on the mean-field evolution of the spinor atomic wavefunction $\boldsymbol{\psi} = (\psi_+, \psi_-)^T$ obeys the system of the Gross-Pitaevskii equations (GPEs), which we here take in the linear form, assuming that the atomic densities are not large enough to introduce appreciable nonlinear effects:

$$i\frac{\partial\psi}{\partial t} = -\frac{1}{2}(\partial_x^2 + \partial_y^2)\psi + i\ \sigma_y\beta_x\partial_x - \sigma_x\beta_y\partial_y\ \psi \\ +[\sigma_z R(x,y) + \alpha y] \tag{1}$$

where $\psi_+$ and $\psi_-$ are the spin-up and down components of the wave function. The SOC Hamiltonian $H \propto \mathbf{p}\cdot\tilde{\mathbf{A}}$, where $\mathbf{p} = -i\hbar\nabla$ refers to the atomic momentum operator and $\tilde{\mathbf{A}} = -\mathbf{i}\sigma_y\beta_x + \mathbf{j}\sigma_x\beta_y$ implies a vector potential acting on the spinor space. The presence of both $\sigma_x$ and $\sigma_y$ terms (via $\beta_y$ and $\beta_x$) ensures that the components of the gauge field $\tilde{A}_x = -\sigma_y\beta_x$ and $\tilde{A}_y = -\sigma_x\beta_y$ do not commute ($[\tilde{A}_x, \tilde{A}_y] = 2i\beta_x\beta_y\sigma_z$), which defines a non-Abelian gauge field. Constants $\beta_{x,y} > 0$ are determined by the interplay of the Rashba SOC and Dresselhaus SOC terms [45, 48–53]. The absence of either $\beta_x$ or $\beta_y$ results in a typical Abelian gauge field ($[\tilde{A}_x, \tilde{A}_y] = 0$). On the other hand, $\beta_x = \beta_y = \beta$ indicates pure Rashba SOC which will reduce the asymmetry of the Bloch dynamics in the present system. Previously, similar gauge terms appeared in models of atomic systems, where they arose from the external coupling of hyperfine states [46]. The essential flexibility sets the atomic systems apart from their solid-state counterparts, where the exact form of SOC is uniquely determined by the lattice symmetry [47, 48]. Experimentally, our model can be implemented with ultracold neutral atoms, such as $^{87}$Rb or $^{40}$K, trapped in a modulated magnetic field. Two internal hyperfine states of atoms (e.g., $F = 1, m_F = 0$ and $F = 1, m_F = \quad 1$ ) coupled by Raman lasers to emulate spin degrees of freedom [39]. The non-Abelian gauge field (Rashba-Dresselhaus SOC) can be synthesized using Raman coupling schemes as pioneered in Refs. [39]. We note that while our focus is on atomic gases, SOC with equal strength of the Rashba-Dresselhaus terms has been implemented in photonic cavities [66, 67]. The universality of the Hamiltonian suggests that our findings for anomalous oscillations can be potentially observed in these photonic settings.

Additionally, a Zeeman lattice [68] may be independently constructed with similar functional profiles, but opposite signs for the two spinor components. The Zeeman lattice is effectively modeled by the potential landscape $R(x,y) = -p\sum_{m,n} \mathrm{Q}(x - x_m, y - y_n)$ with amplitude $p$ and a characteristic width $d$ of lattice sites at points $(x_m, y_n)$ They form a honeycomb grid composed of Gaussian potential wells $\mathrm{Q}(x,y) = \exp[-(x^2 + y^2) / d^2]$, with distance $a$ between adjacent sites. Further, parameter $\alpha$ represents a small potential gradient along the $y$-axis, which is necessary for the occurrence of BO. We assume that the array is periodic along the $y$-axis, with period $Y = \sqrt{3}a$, and it is truncated along the $x$-axis so that it is shaped as a topological insulator with two zigzag edges [a double layer of this structure is displayed in Fig.2 (a)].The lattice is designed to be narrow in the $x$-direction (only 4 lattice units), while long periodic in the $y$-direction. This is because topological edge states exhibit intrinsic spatial localization, which is determined solely by the lattice interface, independent of the overall $x$-width. Increasing the $x$-width would only add insulating bulk regions, which do not alter the physical insights or generality of our results, but imply unnecessarily complicated experimental fabrication and heavier numerical simulations.Eigenmodes of such a truncated lattice are Bloch waves in the form of $\boldsymbol{\psi}(x,y) = \psi(x,y)\exp(iky - i\varepsilon(k)t)$ where $\varepsilon(k)$ is the Bloch eigenenergy, with Bloch momentum $k \in [0, K]$ along the $y$-axis, $K = 2\pi / Y$ being the width of the Brillouin zone, and $\psi(x,y) = \psi(x, y+Y)$ is a periodic spinor wave function localized along the $x$-axis.

## The linear spectrum

The type of SOC, or more precisely the relative magnitudes of the SOC strengths $\beta_x$ and $\beta_y$ is a decisive factor determining whether topological modes can be realized. Consider the unperturbed Hamiltonian $\hat{H}_0$ possessing an additional time-reversal symmetry $T' = \sigma_z T$, which is broken only when both SOC components, $\beta_x\sigma_y k_x$ and $\beta_y\sigma_x k_y$ are present. While the $\beta_y\sigma_x k_y$ term breaks the fundamental time-reversal symmetry $T$, neither it nor the $\beta_x\sigma_y k_x$ term (which preserves $T$) can alone induce topological states even if they open a spectral gap. Only the coexistence of the Rashba and Dresselhaus terms fully breaks $T'$, thereby serving as a necessary condition for the emergence of topological gaps and topological edge states [54].

Fixing $\beta_x$=0.0 while increasing $\beta_y$ or fixing $\beta_y$ =0.0 and increasing $\beta_x$ cannot give rise to topological edge states in the truncated lattice, hence no TBO can be produced, which is clearly visible in Figs.1 (a) and (f). The spectrum of the truncated structure with $\beta_x$=0.0 and increasing $\beta_y$ entails the shift of the Dirac points towards the center, $k$=0, or the edge, $k$=K, of the Brillouin zone. The Dirac points collide at a certain value of $\beta_y$, opening a nontopological gap. With the variation of $\beta_x$, the gap closes and reopens at $\beta_x$~=0.85 when $\beta_y$ =1.5.

The spectra for particular values of $\beta_x$ with fixed $\beta_y$ =1.5, and the same set of value of $\beta_y$ with fixed $\beta_x$ =1.5, are plotted, severally, in Figs.1(a-e) and (f-j), where the black and colored curves represent the bulk and edge modes, respectively. The momentum and energy here is scaled in units of the $\hbar k_L = \dfrac{\sqrt{2}\pi\hbar}{\lambda}$ and $E_L = \dfrac{\hbar^2 {k_L}^2}{2m}$, where $k_L$ is the effective wavevector magnitude corresponding to the Raman laser in the atomic momentum space, and $E_L$ represent Raman recoil energy. The non-Abelian nature of the spectra is indicated by the very different structure of the evolution of these two sets of spectra when $\beta_x \neq \beta_y$, which, nevertheless, become identical at $\beta_x = \beta_y = 1.5$.

The structure of the edge modes denoted by the labeled solid points in Fig. **1** is displayed in Fig. **2**, designated by the same labels. With the same fixed momentum $k = 0.4K$ in the same branch, the increase of $\beta_x$ reduces the localization of the edge states, as seen from the comparison of Figs. **2**(c2) and (e2). Nonetheless, edge modes with $\beta_x = 1.4$ and $\beta_y = 1.5$ are much more compact than their counterparts $\beta_y = 1.4$ and $\beta_x = 1.5$, see Fig. **2**(e2,e3) and (j2,j3). The spin-momentum locking is visible in the mode distributions. The color variation indicates how the spin texture rotates as the momentum $k$ varies, which is the hallmark of the topological phase. The spatial decay of these modes into the bulk dictates the coupling strength between opposite edges during the oscillation cycle. Edge modes denoted by the red point in Fig. **1**(b-d) are selected as the input states, which have a positive group velocity $v(k) = d\varepsilon/dk$ and will be modulated by a wide Gaussian wavepacket with width $w = 30$ and amplitude $a_m = 1$, so that $\psi_{\text{input}} = a_m \exp(-y^2/w^2)\psi(x, y)$, to complete the topological Bloch circle.

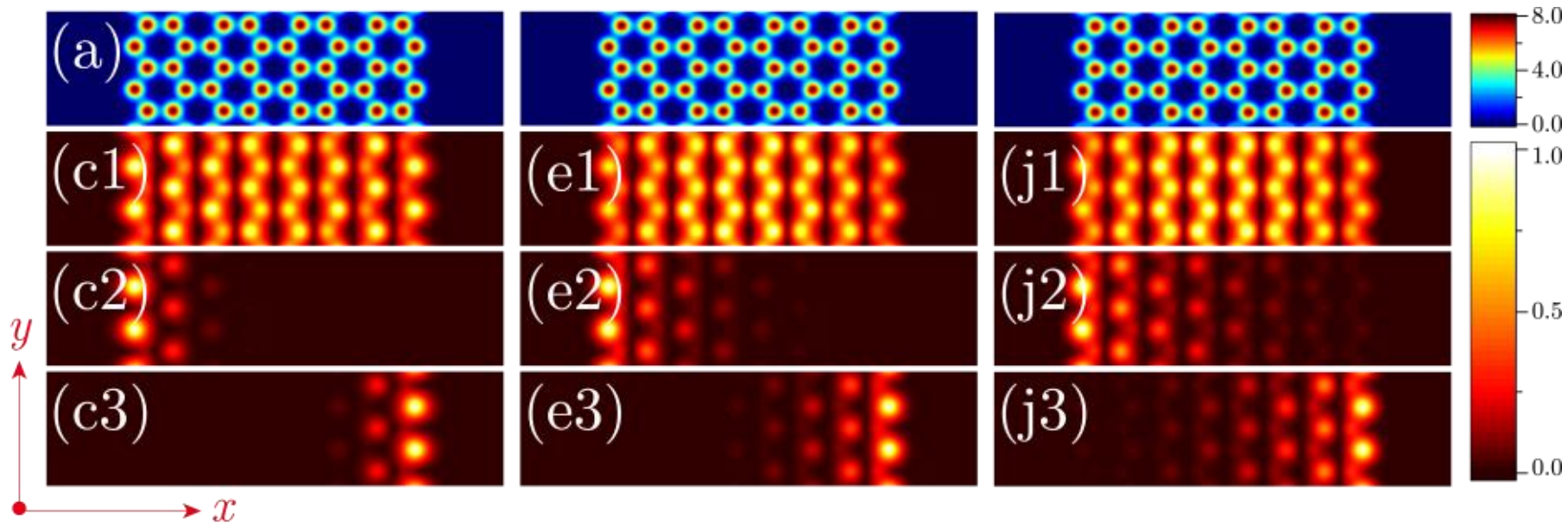

**Fig. 2** (a) The lattice with zigzag-zigzag edges. Panels (c1)-(c3), (e1)-(e3), and (j1)-(j3) displays examples of the distribution of the absolute value of dominating component in modes corresponding, respectively, to points (c1)-(c3), (e1)-(e3), and (j1)-(j3) in the energy-momentum diagrams (c), (e), and (j) in Fig. 1 at $k = 0.01K, 0.4K$, and $0.6K$.

## Discussion

Including a linear gradient in the longitudinal direction of a truncated periodically modulated magnitude field in Eq. (1), one induces an effective gradient force in the $y$-direction, applied to the wavepacket, which triggers TBO, whose dynamics can be detected using in-situ absorption imaging or Time-of-Flight (TOF) expansion [69, 70]. The usual assumption of small variation of the potential within a lattice period is valid for a weak gradient force, which allows one to neglect the effect of the gradient on the profiles of Bloch modes. Here we consider the gradients in the range $\alpha = 0.001$ ~0.01, so that the evolution of the system is adiabatic: one can use the same set of eigenmodes, while, under the action of the force, the Bloch momentum of the wave in our narrow lattice ribbon slowly varies in time, $k(t) = k_0 + \alpha t$, scanning the whole Brillouin zone [1], as discussed below.

The Bloch wave with a broad envelope and momentum $k_0$ moves along the corresponding branch of the dispersion relation, undergoing shape transformations in the coordinate space that correspond to the shift of the wavepacket position in the spectrum from Fig. 1. Since the dependence $\varepsilon(k)$ is periodic, the evolution in the spatial domain is periodic too, if the Landau-Zener tunneling to higher bands [2] is negligible. If one of the modes from the depth of the first or second bands in Fig. 1 is used for the construction of a broad wavepacket, it moves along the corresponding branch of the dispersion relation, remaining in the bulk of the array and exhibiting BO with period $T = \mathrm{K}/\alpha$. The same familiar BO dynamics (without interband transitions and switching between different edges) is observed in nontopological systems, where either SOC is absent, ($\beta_x = 0$) or ($\beta_y = 0$) (recall that there is no topological gap in such a system, hence the wavepacket exciting mode from a certain branch remains in the same band).

To explore how non-Abelian SOC affects ATBO dynamics, we fix $\alpha = 0.001$ (larger values of $\alpha$ may induce Zener tunneling, which would destroy the Bloch oscillation effect). Fig. 3 shows the evolution of the dominant component of the wavefunction under the combined action of Rashba and Dresselhaus SOC terms with unequal coefficients, for fixed $\beta_y = 1.5$ and $\beta_x = 0.86$ (a), 0.96 (b), or 1.06 (c). All initial states carry Bloch momentum $k_0 = 0.4$ K, corresponding to panels **b2,c2,d2** in Figs. 1(b,c,d). A schematic trajectory is overlaid on Fig. 1 to illustrate the edge–bulk–edge motion under the action of $\alpha = 0.001$, where circled numbers from 1 to 5 mark five specific moments at times $t = 0$, 1600, 3000, 4200, 5200, corresponding to $t = 0T, T/4, T/2, 3T/4, T$, respectively.

Initially, the wavepackets move in the positive $y$-direction. Driven by the constant force, they propagate along the excited branch of the dispersion relation and traverse the topological gap, as the state connects two distinct bands. For $\alpha > 0$, the wavepacket thus transforms into a bulk state at the bottom of the second band. This process is generally accompanied by a significant displacement along the $y$-axis and a shift of the wavepacket into the bulk in real space [see Fig. 3(a–c) at a quarter of the BO period, $t \approx 1600$]. Continuing along the dispersion branch near the bottom of the second band, the wavepacket reaches $k = \mathrm{K}$ and, due to the periodicity of the Brillouin zone, reappears at $k = 0$. At this point, the group velocity reverses sign, as clearly shown in Fig. 5.

Further acceleration by the force drives the wavepacket back into the topological gap, until it reaches the points **b3,c3,d3**, which correspond to edge states with negative group velocity localized on the opposite edge [see Fig. 3(a–c) at the TBO half-period, marked by green numbers]. Beyond this point, the wavepacket transitions into a state at the top of the first band, again expanding into the bulk [see Fig. 3(a–c) at three-quarters of the BO period, $\approx 4200$]. After reaching $k = \mathrm{K}$ once more, the wavepacket arrives at positions **b1,c1,d1** and finally returns to its initial location **b2,c2,d2** within the topological gap,i.e., back to the left edge in real space, completing one full ATBO cycle. This dynamical scenario clearly demonstrates that, in stark contrast to nontopological systems or bulk excitations in topological ones, ATBOs involving edge states exhibit a period $T = 2\mathrm{K}/\alpha$, which is twice the conventional Bloch oscillation period. Thus, the wavepacket must traverse the Brillouin zone twice to return to its initial position.

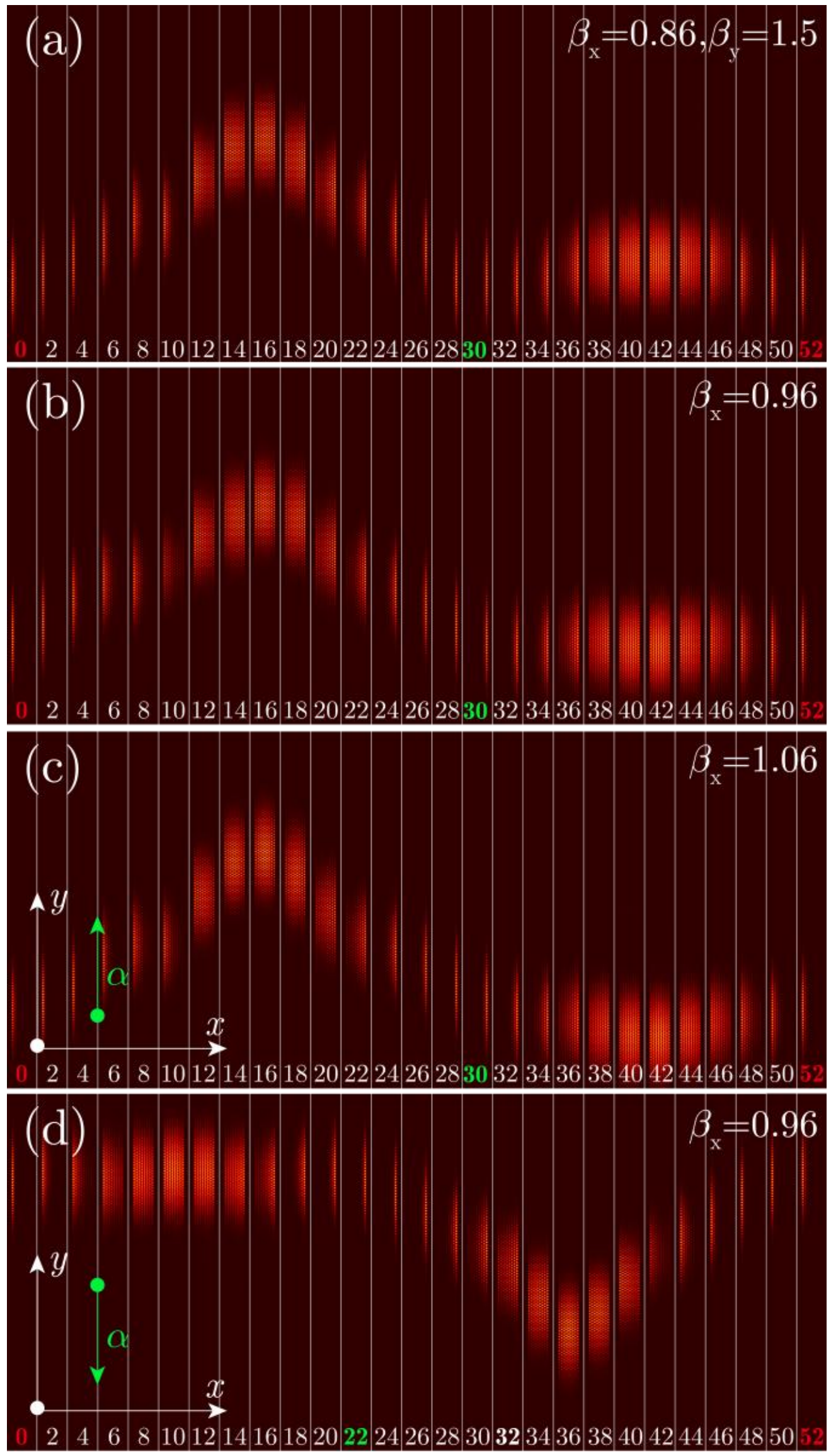

**Fig. 3** Distributions of $|\psi_-|$ at different times, corresponding to the numbers attached to the horizontal axes multiplied by 100. The panels display BO dynamics at fixed $\beta_y = 1.5$, with different values $\beta_x = 0.86$ (a), $\beta_x = 0.96$ (b), and $\beta_x = 1.06$ (c), for the gradient $\alpha = 0.001$. (d) the same as (b), but for a negative gradient, $\alpha = -0.001$. Initially, the topological edge state with momentum $k = 0.4$K and width $w = 30$ is located at the left edge. Green arrows indicate the gradient direction.

In general, the oscillations of the $y$-coordinate of the wavepacket center are out-of-phase between the first and second halves of the cycle. However, the non-Abelian SOC configuration alters this behavior. Figure 3 shows that at $\beta_x = 0.86$ (near the topological transition), the second half of the ATBO becomes in-phase with the first half. As $\beta_x$ increases, the competition between Rashba and Dresselhaus SOC terms suppresses the anomalous dynamics, leading to static localization over an extended duration [Fig. 3(b)], where the "freezing" of TBOs is evident during the second half of the period for $\alpha = 0.001$. This peculiar phenomenology originates from the energy–momentum dispersion shown in Fig. 5. The slowly moving wavepacket gradually restores its shape and returns to the initial position after traversing the Brillouin zone twice, completing one ATBO cycle. A further increase of $\beta_x$ gradually breaks this balance, restoring conventional out-of-phase ATBO dynamics, as clearly seen in Fig. 3(c) and further confirmed by Fig. 4(a).

In Fig. 4(b), we plot the coordinates of the center of mass of the wavepacket in the coordinate space ($y_c$) as functions of time [38, 71], calculated as

$$(x_c, y_c) = U^{-1} \iint (x, y)(|\psi_+|^2 + |\psi_-|^2)dxdy, \tag{2}$$

where $U = \iint(|\psi_+|^2 + |\psi_-|^2)dxdy$ is the total norm of the wave function. Note that we study a relatively narrow topological insulator to reduce the BO temporal period. The period can be drastically reduced by larger gradients $\alpha$, but this may lead to Landau-Zener tunneling. A similar expression can be used in the Fourier domain:

$$(k_x, k_y) = 4\pi F^{-1} \iint (\kappa_x, \kappa_y)(|\tilde{\psi_+}|^2 + |\tilde{\psi_-}|^2)d\kappa_x d\kappa_y,$$

where $F = \iint(|\tilde{\psi_+}|^2 + |\tilde{\psi_-}|^2)d\kappa_x d\kappa_y$ and $\tilde{\psi_+}, \tilde{\psi_-}$ are the Fourier transforms of $\psi_+$ and $\psi_-$. The schematic diagram of the wavepacket motion in the Fourier space is indicated in Fig. 5(b) by arrows across values of the momentum *k/K*, where we plot the dispersion relation with respect to *k/K*.

The center of the wavepacket in the $y$ direction traces its trajectory along the $y$ axis in real space, as shown in Fig. 4(b), where one can clearly observe that the non-Abelian SOC leads to markedly different oscillation amplitudes. At $\beta_x = 0.96$ and $\beta_y = 1.5$, $y_c$ remains nearly constant after the completion of a half-period, and a transition from a positive to a negative $y$-shift occurs as $\beta_x$ increases.

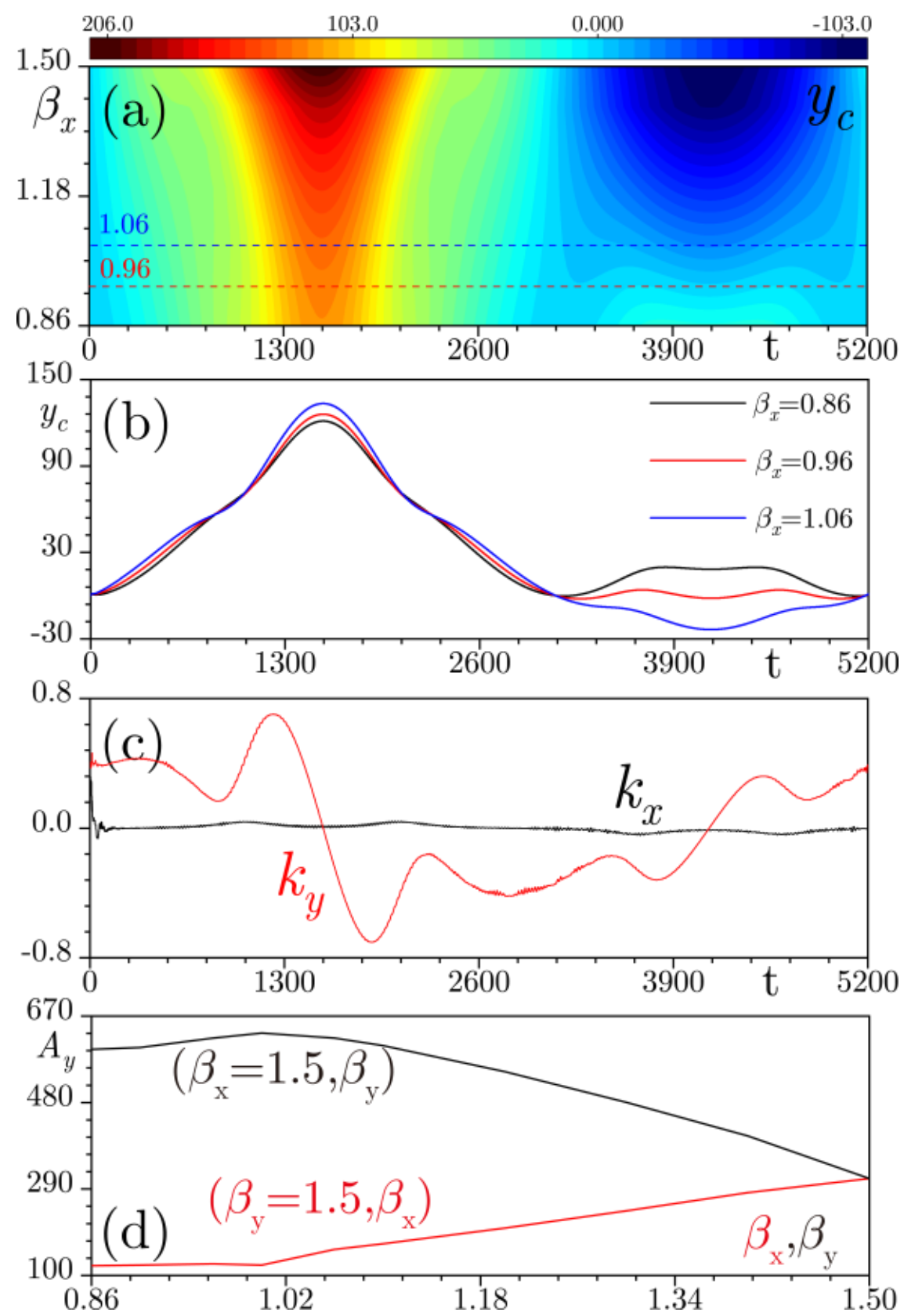

**Fig. 4** (a) $2D$ heat map for $y_c$ dependent on the $\beta_x$ and evolution time $t$. (b) The coordinates of the mass of the wave packet in real space ($x_c$, $y_c$) in the course of TBO corresponding to Fig. 3(a-c). (c) The coordinates of the center of mass of the wave packet in the Fourier domain ($k_x$, $k_y$) as functions of time corresponding to Fig. 3(b). (d) The amplitude of the oscillatory motion in the $y$-direction for the fixed $\beta_{x,y}$ = 1.5, upon the increase of $\beta_{y,x}$, where the black and red curves indicate the dependence on $\beta_y$ or $\beta_x$, respectively.

The $2D$ heat map [see Fig. 4(a)] displays the center-of-mass displacement $y_c$ as a function of evolution time and SOC strength $\beta_x$ at fixed $\beta_y = 1.5$. The red (blue) dashed line indicates the $y_c$ trajectory shown in Fig. 4(b) for $\beta_x = 0.96$ ($\beta_x = 1.06$). This visualization clearly captures the transition from symmetric to asymmetric oscillations, as well as the freezing effect. As $\beta_x$ increases, ATBOs become increasingly symmetric, reaching maximal symmetry in the case of the equal Rashba–Dresselhaus SOC strengths $\beta_x = \beta_y = 1.5$.

It should also be noted that, unlike in non-topological systems, where wavepackets undergo longitudinal oscillations along the force gradient, while remaining confined to a single transverse edge throughout a full Bloch oscillation cycle, in topological systems, the wavepacket must switch between opposite edges via delocalized bulk modes over one complete BO cycle. This is achieved through edge states, which periodically restore their initial positions by transitioning to their counterpropagating counterparts on the opposite boundary. Consequently, TBOs involve both longitudinal oscillations driven by the gradient and transverse edge-to-edge oscillations, resulting in an out-of-phase oscillation pattern of the wavepacket's center $x_c$ (not shown here).

Fig. 4(c) illustrates the periodic motion of the wavepacket and its transverse dynamics in the spectral domain, where the $y$-component exhibits a much larger variation, while the $x$-component shows only slight fluctuations. The $k_{x,y}(t)$ dependencies are perfectly periodic, demonstrating near-complete recovery of the wave packet after a single Bloch oscillation cycle. It should be noted that the integral criterion (3) yields smooth temporal evolutions of $k_{x,y}$, as shown in Fig. 4(c), even when the wave-packet reappears at the opposite edge of the Brillouin zone.

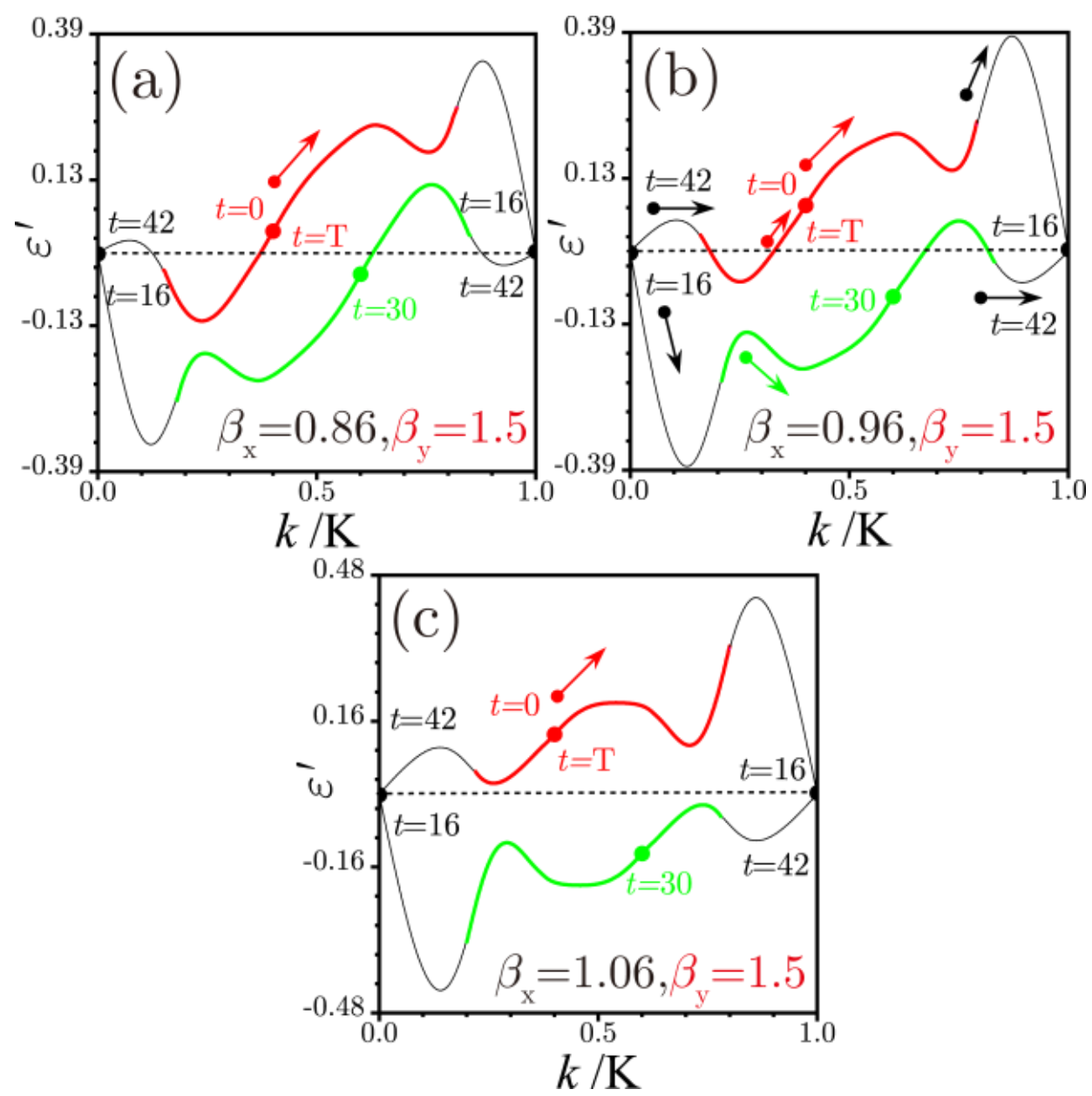

**Fig. 5** The group velocity of the edge modes for $\beta_x$ = 0.86(a), 0.96(b), 1.06(c), and $\beta_y$ = 1.5. Red (green) solid points indicate the moment when the wavepacket is located at the left edge, at $t$ = 0, with momentum $k$ = 0.4K (at the right edge at $t$ = 3200, with momentum $k$ = 0.6K). The colored arrows indicate the motion of the wavepacket across the momentum space.

If the input corresponds to an edge state, inversion of the sign of the gradient does not change the TBOs direction in the coordinate space, whereas the direction of motion in momentum space does reverse. This is because, for $\alpha < 0$, the evolving wavepacket couples to the mode at the top of the first band (rather than to the mode at the bottom of the second band, as occurs for $\alpha > 0$). Consequently, the spatial structure of the wavefunction in the bulk (corresponding to the first and second halves of the ATBOs cycle) differs for opposite signs of the gradient. The evolution of the wavepacket's center, displayed in Figs. 3(b) and (d) for $\alpha = 0.001$ and $\alpha = -0.001$ with $\beta_x = 0.96$, $\beta_y = 1.5$, demonstrates that the sign of $\alpha$ determines whether the wavepacket's motion in the $y$ direction halts during the second or the first half-period of the ATBOs cycle. This result reveals that ATBO dynamics can be controlled not only by the relative SOC strength, but also by the direction of the linear potential gradient.

Figure 4(d) shows the dependence of the $y$-amplitude $A_y$ of ATBOs on the SOC strength $\beta_x$ ($\beta_y$) at fixed $\beta_y = 1.5$ ($\beta_x = 1.5$) for a potential gradient $\alpha = 0.001$. Here, $A_y$ is defined as the difference between the maximum and minimum $y$-positions of the wavepacket during its time evolution. While $A_y$ increases monotonically with $\beta_x$ at fixed $\beta_y = 1.5$ [red curve in Fig. 4(d)], its dependence on $\beta_y$ at fixed $\beta_x = 1.5$ [black curve in Fig. 4(d)] is nonmonotonic. This asymmetry reflects the non-Abelian nature of the underlying dynamics, which does not remain invariant under the exchange $\beta_x \leftrightarrow \beta_y$.

In general, the TBOs amplitude is proportional to the maximal energy difference acquired by the wavepacket as it traverses the Brillouin zone. Typically, this energy difference grows with the SOC strength due to the broadening of the topological gap. This behavior is observed for $\beta_y = 1.5$ and $\beta_x > 1.02$. However, the situation differs for $\beta_x = 1.5$, as $\beta_y$ increases, the topological gap initially widens up to $\beta_y = 1.02$, but then narrows for larger $\beta_y$, resulting in a reduced energy interval scanned by the wavepacket and, consequently, a decrease in $A_y$.

A notable feature occurs at $\beta_x = \beta_y$, where pure Rashba SOC was reached. Despite the fact that the TBOs amplitude always decreases with increasing of SOC strength, the Bloch oscillation period remains independent of it. Moreover, consistent with the adiabatic approximation for wavepacket dynamics in the Brillouin zone under a constant force, $A_y$ scales as $\sim 1/\alpha$.

Figure 5 plots the group velocity along the relevant band branch in momentum space to intuitively illustrate the evolutionary mechanism of ATBOs in both real and Fourier space. The sign reversal of the group velocity at $k = 0$ and $k = K$ reflects the physical transfer of the wavepacket between opposite edges via bulk states, i.e., from the left edge to the right edge or vice versa. The TBO period involves traversal across two energy bands (through gap crossing), which requires integrating the group velocity over a $4\pi$ change in Bloch phase (equivalent to a $2K$ shift in momentum) to complete one full spatial cycle. The color coding in this figure follows the same convention as the band structures shown in Fig. 1.

Take $\alpha = 0.001$ as an example, following the arrows in Fig. 5(b), the wavepacket starting from the red point ($k = 0.4K$, $t = 0$) moves upward along the red branch, penetrates into the bulk band denoted by the black curve, reaches the momentum value $k = K$, and then reappears at $k = 0K$ due to the periodicity of the Brillouin zone. The wavepacket subsequently acquires a negative group velocity, moves downward, and arrives at $k = 0.6K$, marked by the solid green point. After a short propagation time, it enters a region where the group velocity is nearly zero, leading to an apparent freezing of ATBOs in the $y$ direction. This frozen state persists until the wavepacket restores its original shape and returns to $k = 0.4K$, completing a full ATBO cycle. This demonstrates that the wavepacket's motion can be effectively halted during the second half of the ATBO cycle.

In contrast, when $\alpha = -0.001$ is applied to the Zeeman lattice, the wavepacket remains quiescent during the first half of the ATBO cycle and only develops oscillations in the second half. Similar behavior is observed for other parameter sets, such as in Figs. 5(a) and (c), corresponding to $\beta_x = 0.86$ (a) and 1.06 (c) at fixed $\beta_y = 1.5$. However, one must pay attention to the sign and magnitude of the dispersion after the wavepacket reaches the opposite turning point at $k = 0.6K$, which marks the completion of the half-period of the TBO cycle. These results confirm that the dispersion relation fully governs the dynamics induced by the non-Abelian gauge field.

## Conclusion

In this work, we investigate topological Bloch oscillations (TBO) in non-Abelian gauge fields, driven by unequal-strength Rashba-Dresselhaus spin-orbit coupling (SOC) under a linear force aligned with a modulated Zeeman field, which can be experimentally realized by integrating a conventional lattice with a magnetic field. The central finding is that the relative weight of Rashba-Dresselhaus SOC terms directly engineers the band structure, giving rise to unique SOC-controlled ATBO modes. We verified that the dynamics of the wavepacket's center of mass are closely related to the Rashba-Dresselhaus SOC and its corresponding group velocity variation during the first and second halves of the ATBOs cycle. Distinct from Abelian systems, these modes originate from the non-commutativity of non-Abelian gauge potentials and are associated with spin-momentum locking effects. This work establishes a foundational framework for mean-field dynamics in non-Abelian gauge fields with SOC, offering tunable control over atomic transport.

As a development of the work, it may be interesting to add nonlinearity to the system and explore the respective effects, such as the formation of solitons and vortices. Future directions can further include investigating many-body interactions in non-Abelian SOC environments, exploring higher-dimensional topological bands, and the realization of the non-Abelian Aharonov-Bohm interference, all of which can be built on the linear regime reported here, expanding the scope of quantum simulation for complex non-Abelian phenomena. Our findings advance the fundamental understanding of the interplay between the non-Abelian SOC and TBOs, and provide suggestions for the experimental realization of SOC-controlled atomic transport.

## Acknowledgements

Chunyan Li acknowledges support from the National Natural Science Foundation of China (NSFC) (Grant No. 12574365), Natural Science Basic Research Program in Shaanxi Province of China (Grant No. 2025JC-YBMS-011), and Fundamental Research Funds for the Central Universities (Grant No. ZYTS25122). Ce Shang is supported by Project Nos. E4BA270100, E4Z127010F, E4Z6270100, E53327020D of the Chinese Academy of Sciences.

## Data availability statement

The datasets generated and analyzed in the current study are available from the corresponding author upon reasonable request.

## CRediT authorship contribution statement

All authors contribute greatly to the work.

## Competing interests

The authors declare no competing interests.

* chunyanli@xidian.edu.cn
† shangce@aircas.ac.cn